\documentstyle[amscd,amssymb,verbatim,12pt]{amsart}
\def\dspace{\baselineskip=0.3 in}
\begin{document}
\dspace
\title[CAN PHANTOM-DOMINATED UNIVERSE ...]{CAN PHANTOM-DOMINATED
  UNIVERSE DECELERATE ALSO IN FUTURE ?}

\author{\bf S.K.Srivastava}
{ }
\maketitle
\centerline{ Department of Mathematics,}
 \centerline{ North Eastrn Hill University,}
 \centerline{  Shillong-793022, India}
\centerline{ srivastava@@.nehu.ac.in; sushilsrivastava52@@gmail.com}

\vspace{0.5cm}

\centerline{\bf Abstract}

Here Randall-Sundrum brane-gravity models of the homogeneous and flat
universe, dominated by 
phantom fluid, is considered. It is noted that, in future, brane-gravity
corrections will
effect  the behaviour of phantom fluid in RS-II model( where
brane-tension $\lambda$ is negative) drastically. It is interesting to see that, phantom
fluid violates the weak energy condition (WEC) till energy density $\rho <
\lambda,$ but when phantom energy density $\rho$
grows more, the effective equation of state does not violate WEC. Moreover,
with increasing phantom energy density, a stage comes when even strong energy
condition is not violated due to effect of these corrections. Also,
expansion stops, when $\rho = 2 \lambda.$ As a cosequence, RS-II phantom
universe accelerates upto a finite time explaining the present cosmic
acceleration , but it decelerates later. Thus,  RS-II phantom
universe is singularity-free.  In the case of RS-I model, where
$\lambda$ 
is positive, characterisics of phantom fluid is not suppressed by
brane-corrections and phantom universe accelerates ending up in big-rip
singularity in finite future time.

\vspace{2cm}

\centerline{\bf 1. Introduction}
\smallskip

Astrophysical observations made, by  end of the last century and beginning of
this century, have conclusive evidence for late cosmic acceleration \cite{sp,
  ag}. It is driven by an unknown fluid violating strong energy condition
(SEC) such that $1 + 3 w < 0$ with $w$ being the ratio of pressure density $p$ and
energy density $\rho$. This exotic fluid is known as dark energy (DE).  In 2002-2003,
Caldwell  argued that experimental observations favor the case $w < -1$ more
than the case $w > -1$
violating the weak energy condition (WEC) too. DE, obeying $w < -1$, is dubbed as phantom \cite{cal}. Using
general 
relativity (GR)-based Friedmann equation, giving cosmic dynamics, it is found
that phantom-dominated era of the universe accelerates, but ends up in big-rip
singularity in a finite future time. One can refer to \cite{ejc}for
detailed review. Later on, in \cite{vs}, using brane-gravity(BG),
singularity-free phantom driven cosmic acceleration was obtained. After some time, in \cite{sks05}, it is
shown that GR-based theory 
also gives accelerating phantom universe if the phantum fluid behaves as
barotropic fluid and generalized Chaplygin gas simultaneously.

In the race of probe for viable cosmological models satisfying
observational constraints and explaining present cosmic acceleration, brane-gravity was
also drawn into service and brane-cosmology was developed. A review on
brane-gravity and its various applications with special attention to cosmology
is available in\cite{var, rm,   pbx, cs}. The brane-gravity stems from low energy string theory.
After development of M-theory , bringing different string-theories under one
umbrella, Ho\v{r}ova and Witten proposed that 11-dimensional supergravity,
which is a supermembrane theory, can be obtained as low-energy limit of 11-dimensional
M-theory. They discussed that it can be done  on a particular orbifold $R^{10}
\otimes S^1/Z_2$, where $R^{10}$ is the 10-dimensional space-time and
$S^1/Z_2$ is the 1-dimensional space having  $ x^{11} \leftrightarrows -
x^{11}$ symmetry \cite{ho}. When the six extra-dimensions on $(1 + 9)$-branes are
compactified on very small scale close to fundamental scale, their effect
is realized on $(1 + 3)$-dimensional brane located at ends of $ S^1/Z_2$.  Thus,  Ho\v{r}ova - Witten solution provided an effective 5-dimensional model
where extra dimension could be large relative to the fundamental scale in
contrast to Kaluza-Klein approach where extra dimension is very small
\cite{app, skp}. In the 5-dimensional model, situation is simplified by the
idea that matter fields are confined to 4-dimensinal space-time, called
3-brane, which is our observable universe and major portion of gravity is
confined to 5-dimensional ``bulk''. The pioneering work, in this direction,
was done by L.Randall and R.Sundrum in their seminal paper to solve
the ``hierarchy problem'' by a {\em warped} or curved dimension showing that
fundamental scale can be brought down from the Planck scale to 100 GeV. Thus,
Randall-Sundrum approach brings the theory to low scales upto 100 GeV, which is
the electroweak scale( so far results could be verified experimentally upto
this scale only). In this model, extra-dimension is
large having $(1 + 3)$-branes at its ends. These branes are $Z_2$-symmetric (have
mirror symmetry) and have tension to counter the negative cosmological
constant in the ``bulk'', which is ${\rm AdS}_5$. The model, having {\em two}
$(1 + 3)$-branes at the ends of the orbifold $S^1/Z_2$, is known as RS-I model \cite{rs1}.

In another paper, in the same year, these authors proposed another brane-model
as an alternative to compactification . In this model, there is only {\em one}
$(1 + 3)$-brane at one end of the extra-dimension and the other end tends to
infinity. This model is known as RS-II model \cite{rs2}. Thus, Ho\v{r}ova -
Witten solution and RS- theory yielded
brane-gravity originating from low-energy string theory, which also explains why gravity is weak in the observable
universe. In case the extra-dimension is time-dependent, brane-gravity induced
Friedmann equation (FE)(giving dynamics of the universe) contains a correction
term $\rho^2/2\lambda$ with $\lambda$ being called as brane-tension. In RS-I
model, $\lambda$ is positive, whereas in RS-II
model, $\lambda$ is negative.

So, apart from general relativity (GR)-based models and $f(R)-$ models,
brane-gravity (BG)-based cosmological models were also tried upon to explain
acceleration in the present and late universe. In particular, RS-II model got much attention due to its simple
and rich conceptual base \cite{pb00, pbd00, cg, cgs, mwb, cll, cal, apt, nopl, sksgrg}.
Here also, we address to RS-II model. In this case, we have energy density
terms in FE as $\rho (1 - \rho/2|\lambda|)$ due to $\lambda < 0$. In
\cite{vs}, it is shown that phantom-dominated model of the universe based on
RS-II model of brane-gravity accelerates till $\rho = 2 |\lambda|$ giving late
transient acceleration.

As mentioned above, in the RS-II model of brane-gravity,
Friedmann equation gets modified by a correction term $- {4\pi G
  \rho^2}/{3\lambda}$\cite{var, rm,   pbx, cs}. The present density of dark
energy is found to be $0.73 
\rho^0_{\rm cr}$ with present critical energy density $ \rho^0_{\rm cr} = 2.5
\times 10^{-47} {\rm GeV}^4$\cite{sp, sp03}. If the late universe is dominated
by phantom , present phantom energy density is $0.73
\rho^0_{\rm cr}$. The brane-tension, in RS-models, are obtained as $\lambda = 48
\pi/ M_P^2 k_5^2$ with $k_5^2 = 8 \pi G_5 =  8 \pi/M_P^2 l$ and $l$ being the
length of the extra-dimsion of the 5-dimensional bulk\cite{rm,   pbx, cs}. Here, we take  $k_5^2 = 1 {\rm GeV}^{-6}$ like the ref.\cite{pbx}, so $\lambda = 48
\pi/ M_P^2 = 6.03 \times 10^{10} \rho^0_{\rm cr} $. Thus, it is found that, in
the present universe, brane-correction term  $- {4\pi G 
  \rho^2}/{3\lambda}$ is ineffective and it may continue so if energy density
decreases with the expansion of the future universe. But, phantom fluid is {\em different} from
other fluids, in the sense that, contrary to other fluids, its energy density 
increases with expansion of the universe due
to EOS parameter $w < -1.$ So, in future, $\rho$ will
increase rapidly due to accelerated expansion being comparable to
$\lambda$. As a result, brane-corrections will be effective in the future universe
being addressed here. This crucial point is elaborated more explicitly in 
following sections. 

The present paper aims to explore some other
aspects of the universe based on RS-II model of brane-gravity, when it is dominated by phantom fluid. In what
follows, it is found that phantom characteristic ( violation of WEC) of the
dominating fluid is 
effective in RS-II model so long as $\rho < 3|1 + w|\lambda/(1
+ 3 |1 + w|) $ and universe
accelerates. $\rho$ of phantom dark energy grows with cosmic expansion. As it
grows to $\rho \ge 3|1 + w|\lambda/(1
+ 3 |1 + w|),$ WEC is not violated but SEC is violated so long
as $\rho < {(2 + 3|1 + w|)}/{(1 + 3|1 +w|)}\lambda$ and acceleration
continues. When $\rho \ge {(2 + 3|1 +  w|)}/{(1 + 3|1 +w|)}\lambda$, SEC is
not violated and even dark energy property of the fluid is supressed by
brane-gravity effects. As a result, universe decelerates. These are 
some new surprises from brane-world cosmology in addition to many surprises
mentioned in \cite{var05} .

Here, natural units ${\hbar} = c = 1$ are used,where $\hbar$ and $c$ have
their standard meaning.

\bigskip

\centerline{\bf 2. Strong and weak energy conditions for phantom fluid}

\centerline{\bf in RS-II model of the universe}

\smallskip

Observations support homogeneous and isotropic model for the late universe \cite{ad},
given by the anastaz
$$ ds^2 = dt^2 - a^2(t) [dx^2 + dy^2 + dz^2], \eqno(1)$$
where $a(t)$ is the scale factor.
 
In this space-time, modified Friedmann equation in RS-II model is given as
$$H^2 = \Big(\frac{\dot a}{a} \Big)^2 = \frac{8\pi G}{3} \rho\Big[1 -
\frac{\rho}{2\lambda}\Big], \eqno(2a)$$
with $G = M_P^{-2}$ ($M_P = 10^{19} {\rm GeV}$ being the Planck mass in
natural units), $\rho$  being energy density and  brane-tension given as
$$\lambda = \frac{48 pi}{ M_P^2} = 6.03 \times 10^{10} \rho^0_{\rm cr}
\eqno(2b)$$ 
on taking $k_5^2 = 1$ as mentioned above \cite{rm, pbx}. Here $\rho^0_{\rm cr}
= 2.5 \times 10^{-47} {\rm GeV}^4$ is the critical energy density in the
present universe.

As mentioned above, observations support dominance of phantom fluid in the late
universe \cite{cal}. So, the equation of state (EOS) of the dominating phantom fluid is taken as 
$$ p = w \rho \eqno(3)$$
with $w < - 1.$

The conservation equation is given by
$${\dot \rho} + 3 H (\rho + p) = 0 \eqno(4a)$$
as it does not contain brane-correction terms \cite{rm} .

 So, for phantom fluid, (4a) looks like
$$ {\dot \rho} -  3\frac{\dot a}{a} \rho |1 + w| = 0  $$
integrating to
$$ \rho  = 0.73 \rho^0_{\rm cr} a^{3|1 + w|}, \eqno(4b)$$
where the present scale factor $a_0$ is normalized as $a_0 = 1.$ 

From (2b) and (4b), it is found that, in the present universe, $\rho^0/
\lambda \simeq 1.2 \times 10^{-11} \approx 0.$ So, brane-correction is not
effective in the present universe. But, as phantom energy density will increase
with growing $a(t)$ given by (4b), this situation will improve in future,  when scale factor
will grow  sufficiently  such that 
$$ a(t)/a_0 \gtrsim [8.26 \times 10^{10}]^{1/3|1 + w|}  \eqno(4c)$$
with $a_0 = a(t_0)$ being the present scale factor and $t_0$ being the present
time. When it will happen so,  $\rho/\lambda \gtrsim 1$ and brane-correction terms
will be  effective in the dynamics of the universe and gradually
brane-corrections will grow stronger. 

Here, {\em three} situations are obtained. As mentioned above. in
the present universe, brane-corrections are not effective. This situation will
continue till $\rho$ grows sufficiently. During this period WEC is violated
and phantom universe will super-accelerate (it is justified at the end of this
section). As for $\rho << 2\lambda$,
universe will suoer-accelerate in future and $\rho$ will grow with $a(t)$, it
is reasonable to believe  $\rho/\lambda \gtrsim 1$ to happen in future due to
rapid increase in $a(t)$ caused by super-acceleration. Increase in $\rho$ will
still continue with growing 
$a(t)$. On further increase in $\rho$, upto certain value,
brane-corrections will be effective and only SEC
will be violated. As a consequence, acceleration of the phantom universe will
become comparatively slow. It means that, in this situation, phantom universe
will accelerate, but it will not super-accelerate. This is the intermadiate
state. When $\rho$ will increase more, none of SEC and WEC will be violated
due to strong effect of brane-corrections and phantom universe will
decelerate.  Acceleration and super-acceleration manifest anti-gravity effect
of dark energy. Here, it is found that that brane-corrections, in RS-II model,
counter anti-gravity effect of phantom dark energy. In the case of
quintessence, $\rho$ decreases with expansion of the universe due to $w > -
1$, so brane-corrections can not be effective in present and future
quintessence universe in RS-II model. In what follows, a detailed analysis is given

From (2a) and (4a), it is obtained that
$$\frac{\ddot a}{a} = - \frac{4\pi G}{3} \Big[3(\rho + p)\Big[1 -
\frac{\rho}{\lambda}\Big] - 2 \rho\Big\{1 -
\frac{\rho}{2\lambda}\Big\} \Big], \eqno(5)$$
This equation was obtained earlier in \cite{nopl, sksgrg}.

In the GR-based theory, (5) looks like
$$\frac{\ddot a}{a} = - \frac{4\pi G}{3} [\rho + 3 P] . \eqno(6)$$

Comparing (5) and (6), the effective pressure density $P$ is given by
$$ \rho + 3 P = 3(\rho + p)\Big[1 -
\frac{\rho}{\lambda}\Big] - 2 \rho\Big[1 -
\frac{\rho}{2\lambda}\Big] . \eqno(7)$$

Using the equation of state (3) for phantom fluid, effective equation of state (EEOS)
with brane-gravity corrections is obtained as
$$ P = - \rho - |1 + w| \rho + \frac{\rho^2}{3\lambda}( 1 + 3 |1 + w|),  \eqno(8)$$
where $(1 + w) = - |1 + w|$ for phantom fluid.

(8) yields
$$ \rho + P = - \rho \Big[ |1 + w| - \frac{\rho^2}{3\lambda}(1 + 3 |1 + w|)
\Big].  \eqno(9)$$
This equation shows that WEC is violated when ${\rho}/{\lambda} < 3|1 + w|/(1
+ 3 |1 + w|) $, $\rho + P = 0$ for ${\rho}/{\lambda} = 3|1 + w|/(1
+ 3 |1 + w|) $ and $\rho + P > 0$ for ${\rho}/{\lambda} > 3|1 + w|/(1
+ 3 |1 + w|) $. 

 It is interesting to note from
(4b) and (9)  that phantom
fluid, dominating the universe, behaves effectively as
phantom  ie. it violates WEC  till ${\rho}/{\lambda} < 3|1 + w|/(1 + 3 |1 + w|),$
but phantom characterisic to violate WEC is suppressed by
brane-gravity effects as $\rho$ 
increases and it obeys the inequality ${\rho}/{\lambda} > 3|1 + w|/(1
+ 3 |1 + w|) $.

Further, using (8), it is also obtained that
$$\rho + 3 P = - \rho \Big[2 + 3|1 + w| - \frac{\rho^2}{\lambda}(1 + 3 |1 + w|)
\Big].  \eqno(10a)$$
(10a) shows that SEC is violated when ${\rho}/{\lambda} < (2 + 3|1 + w|)/(1
+ 3 |1 + w|) $, $\rho + 3 P = 0$ for ${\rho}/{\lambda} = (2 + 3|1 + w|)/(1
+ 3 |1 + w|)$ and $\rho + 3 P > 0$ for ${\rho}/{\lambda} > (2 + 3|1 + w|)/(1
+ 3 |1 + w|)$. 

Thus, it is obtained that (i) WEC is violated for ${\rho}/{\lambda} < 3|1 +
w|/(1 + 3 |1 + w|) $, (ii) for $3|1 + w|/(1 + 3 |1 + w|) \le {\rho}/{\lambda}
< (2 + 3|1 + w|)/(1 + 3 |1 + w|) $ WEC is not violated, but SEC is violated
and (iii) for ${\rho}/{\lambda} > (2 + 3|1 + w|)/(1
+ 3 |1 + w|)$ neither of the two conditions is violated. Also it is
interesting to 
note that brane-gravity corrections cause {\em phantom divide} when
${\rho}/{\lambda} =  3|1 + w|/(1 + 3 |1 + w|)$ \cite{sks06}. 

Using these results in (5), it is obtained that universe will accelerate
 till ${\rho}/{\lambda}< (2 + 3|1 + w|)/(1 + 3 |1 + w|) $ as ${\ddot a} >
 0$. At ${\rho}/{\lambda} = (2 + 3|1 + w|)/(1 + 3 |1 + w|), {\ddot a} = 0$ and
 ${\rho}/{\lambda} > (2 + 3|1 + w|)/(1 + 3 |1 + w|), {\ddot a} < 0.$ It shows
 that a transition from acceleration to deceleration will take place when
 ${\rho}/{\lambda} = (2 + 3|1 + w|)/(1 + 3 |1 + w|).$

In a flat universe, combined analysis of results from WMAP3 and the Supernova
Legacy Survey (SNLS) yields $ w = - 0.97^{0.07}_{-0.09}$. If flat universe is
not considered a priori, WMAP3 with large structure and supernova data yield $
w = - 1.06^{0.13}_{-0.08}$ \cite{dn06}. Here, for an example, we take a 
value for $ w $ as $w = - 1.06$ common to both observational results.

Using these values of $\lambda$ and $w$, the {\em effective phantom divide} is
obtained at 
$${\rho} =  1.94\times 10^{10} \rho^0_{\rm cr}, \eqno(10b)$$  
transition from acceleration to deceleration is obtained when
$${\rho} =  10^{11} \rho^0_{\rm cr}, \eqno(10c)$$ 
and it is found that universe will expand till
$${\rho} = 1.19\times 10^{11} \rho^0_{\rm cr}. \eqno(10d)$$

According 16 Type supernova observations \cite{ag}, acceleration in the late
universe started at red shift $z_* \simeq 0.46$ and a transition from
deceleration to acceleration took place. So, the scale factor at this
red-shift is obtained from
$$\frac{a_0}{a_*} = 1 + z_* = 1.46. \eqno(10e)$$
 
From (4b), we have 
 $$ \rho = 0.73 \rho^0_{\rm cr} (a/a_0)^{0.18} \eqno(10f)$$
 using $w = - 1.06$. So,
$$ \rho_* = 0.73 \rho^0_{\rm cr} (a_*/a_0)^{0.18} = 0.68 \rho^0_{\rm   cr}. \eqno(10g)$$

When WEC is violated, from  (6), it is obtained that ${\ddot a}/a > 8\pi G
\rho/3$ and $0 <{\ddot a}/a < 8\pi G \rho/3 ,$ when only SEC is
violated. Thus, ${\ddot a}/a \rho$ is greater during violation of WEC than
${\ddot a}/a \rho$ during violation of SEC only. It shows supera-acceleration,
when WEC is violated and acceleration , when SEC is violated.
So, using above numerical values, it is obtained that universe will
 super-accelerate when $ \rho_* < \rho < 9.198\times 10^{9} \rho^0_{\rm cr}$
 as EEOS for phantom fluid violates  WEC during this range of energy
 density. The universe will accelerate when $9.198\times 10^{9} \rho^0_{\rm
 cr}   < \rho < 1.11\times 10^{11} \rho^0_{\rm cr}$ as EEOS for phantom fluid does not
 violate  WEC, but violates SEC ie. phantom fluid behaves effectively as
 quintessence during this range of $\rho$. The universe will decelerate when
 $1.11\times 10^{11} \rho^0_{\rm cr} < \rho < 2.06 \times 10^{11} \rho^0_{\rm cr}$ as
 during this range of $\rho$, neither SEC nor WEC is violated by EEOS of
 phantom fluid giving ${\ddot a} < 0.$

\bigskip

\centerline{\bf 3. Acceleration and deceleration of the universe}

\smallskip

So far, we have obtained different conditions for changes in the behaviour of
phantom fluid dominating the RS-II model-based  universe due to brane-gravity
corrections. In what follows, we derive scale factor $a(t)$ solving Friedmann
equation (2a) and conservation equation (4a). It helps to find time period
during which WEC and SEC are violated and time period, during which, these are
not violated.

Connecting (2a) and (4a) and using $(1 + w) = - |1 + w|$ for phantom fluid, it is obtained that
$${\dot \rho} - 3|1 + w|\rho \sqrt{\frac{8\pi G}{3} \rho\Big[1 -
\frac{\rho}{2\lambda}\Big]} = 0 .\eqno(11)$$
(11) is integrated to
$$ \rho = \Big[\frac{1}{2\lambda} + \Big\{\sqrt{\frac{1}{\rho_*} -
  \frac{1}{2\lambda}} - \sqrt{6\pi G}|1 + w|(t - t_*) \Big\}^2 \Big]^{-1} ,
  \eqno(12)$$
where $\rho_* < 2\lambda$ is the phantom dark energy density and
  $t_*$ is the time, when late universe began to accelerate at red-shift $z_*
  \simeq 0.46$ given above. It shows that phantom energy density increases
  with time as it is obtained in GR-based theory also.

From (2) and (12), it is obtained that
$$H^2 = \frac{8\pi G}{3}\frac{\Big\{\sqrt{\frac{1}{\rho_*} - 
  \frac{1}{2\lambda}} - \sqrt{6\pi G}|1 + w|(t - t_*) \Big\}^2}{\Big[\frac{1}{2\lambda} + \Big\{\sqrt{\frac{1}{\rho_*} -
  \frac{1}{2\lambda}} - \sqrt{6\pi G}|1 + w|(t - t_*) \Big\}^2 \Big]^2}. \eqno(13)$$

(13) yields the solution
$$ a(t) = a_* \rho_0^{-1/3|1 + w|}\Big[\frac{1}{2\lambda} +\Big\{\sqrt{\frac{1}{\rho_*} -
  \frac{1}{2\lambda}} - \sqrt{6\pi G}|1 + w|(t - t_*) \Big\}^2\Big]^{-1/3|1 + w|} \eqno(14)$$
with $a_* = a(t_0)/(1 + z_*) = 0.68 a_0$ using (10e). It is evident from (11) and (13) that this model does not
  encounter any finite time future singularity.Using $\rho = 2\lambda$ in
  (11), it is obtained that phantom-era ends at time
$$t_e = t_* + \frac{1}{\sqrt{6\pi G}|1 + w|}\sqrt{\frac{1}{\rho_*} -
  \frac{1}{2\lambda}}.  \eqno(15)$$

The phantom-dominated model, discussed here, expands between the energy
density interval $\rho_* \le \rho \le 2\lambda.$ As noted above, this interval
has two crucial points (i) $\rho = \lambda \Big[3 |1 + w |/(1 + 3 |1 +
w|)\Big]$ and (ii) $\rho = \lambda \Big[(2 + 3 |1 + w |)/(1 + 3 |1 + w|)\Big].$

Using (12), it is found that the first point, giving {\em effective phantom divide},
is reached at time 
$$ t_1 = t_* + \frac{1}{|1 + w|\sqrt{6\pi G \rho_*}} \Big[\sqrt{1 -
    \frac{\rho_*}{2\lambda}} - \sqrt{\frac{(2 + 3 |1 + w|) \rho_*}{6\lambda |1 + w|}}
    \Big]  \eqno(16)$$
and the second point, giving transition from acceleration to deceleration, is
    reached at time 
$$ t_2 = t_* + \frac{1}{|1 + w|\sqrt{6\pi G \rho_*}} \Big[\sqrt{1 -
    \frac{\rho_*}{2\lambda}} - \sqrt{\frac{(3 |1 + w|) \rho_*}{2\lambda (2 + 3|1 +
    w|)}} \Big].  \eqno(17)$$

These results are supported by (14) giving ${\ddot a } > 0,$ when 
$$ t < t_* + \frac{1}{|1 + w|\sqrt{6\pi G \rho_*}} \Big[\sqrt{1 -
    \frac{\rho_*}{2\lambda}} - \sqrt{\frac{(3 |1 + w|) \rho_*}{2\lambda (2 +
    3|1 +  w|)}} \Big].  $$

As $t_1 > t_0$ and $t_2 > t_0$, (16) and (17) yield a constraint on
brane-tension as
$$ \lambda > \rho_0 \Big[\frac{(1 + 3 |1 + w|)}{3 |1 + w|}\Big]. \eqno(18)$$

Thus, it is found that, in this model, dominating phantom fluid behaves as phantom due
to $(\rho + P)$ being negative during the time interval $t_0 < t < t_1$. It
behaves effectively like quintessence when $t_1 < t <
t_2$ as $(\rho + P) > 0$ and $(\rho + 3 P) < 0$  during this time
interval. So, universe accelerates during the time interval $t_0 < t <
t_2$. During interval $t_2 < t < t_e,$ the effective equation of state obeys
the condition $(\rho + 3 P) > 0,$ so universe decelerates. Thus $t = t_2$
gives the transition time from acceleration to deceleration.

From (15) and (17), it is found that if $\lambda$ is very large, $t_2 \approx t_e$ and
deceleration period $t_e - t_2 $ vanishes. If $\lambda$ is small ,
deceleration period $t_e - t_2 $ is significant. It shows that $\lambda$ has a
very crucial role.

Using value of $\lambda$ given by (2b) and $ w = - 1.16$ in (15) and (16),
it is obtained that
$$ t_1 =  t_e  - 393.19 {\rm kyr}, $$
where present age of the universe is $ t_0  = 13.7 {\rm Gyr} = 6.6 \times
10^{41} {\rm GeV}^4$. This result shows that effective phantom divide will be
obtained $393.19 {\rm kyr}$ before $t_e = 6.1 t_0 = 83.57 {\rm Gyr}$. 

Similarly, using these numerical values in (15) and (17), we have
$$ t_2 =  t_e  - 88.96 {\rm kyr}. $$
This result shows that transition from acceleration to deceleration will take
place $ 88.96 {\rm kyr}$ before $t_e.$

Using $a = a_*$, given by (10f), at $t = t_*$, (14) yields
$$ t_* = 0.84 t_0 = 11.47 {\rm Gyr}, $$
where above values of $\lambda$ and $w$ are used. Thus, it is obtained that
universe will super-accelerate during the time period $11.47 {\rm Gyr} < t <
t_e - 393.19 {\rm kyr}$ as EEOS violates WEC during this period, will acceletrate during the period $ t_e  - 393.19
{\rm kyr} < t <  t_e  - 88.96 {\rm kyr} $ as EEOS violates SEC, but not WEC during this period and will decelerate during the
period $t_e  - 88.96 {\rm kyr} < t <  t_e$ as neither WEC nor SEC is violated
duting this period.

In case of RSI model $\lambda$ is positive, so (2) looks like
$$ \Big(\frac{\dot a}{a} \Big)^2 = \frac{8\pi G}{3} \rho\Big[1 +
\frac{\rho}{2\lambda}\Big], \eqno(19)$$

In this case, we have
$$ \rho = \Big[-\frac{1}{2\lambda} + \Big\{\sqrt{\frac{1}{\rho_0} +
  \frac{1}{2\lambda}} - \sqrt{6\pi G}|1 + w|(t - t_0) \Big\}^2 \Big]^{-1} ,
  \eqno(20)$$
$$ a(t) = a_0 \rho_0^{-1/3|1 + w|}\Big[\frac{1}{2\lambda} +\Big\{\sqrt{\frac{1}{\rho_0} +
  \frac{1}{2\lambda}} - \sqrt{6\pi G}|1 + w|(t - t_0) \Big\}\Big]^{-1/3|1 + w|} \eqno(21)$$
and
$$\frac{\ddot a}{a} = \frac{ 4\pi G}{3}\rho \Big[(2 + 3|1 + w|) + (1 + 3|1 +
w|)\frac{\rho}{\lambda}\Big] > 0.  \eqno(22)$$

(16) and (17) show big-rip singularity at time
$$ t_s = \frac{1}{\sqrt{6\pi G}|1 + w|} \Big[\sqrt{\frac{1}{\rho_0} +
  \frac{1}{2\lambda}}\Big] \eqno(23)$$
in RSI model as $\rho \to \infty, p \to \infty$, when $t \to t_s$ as it is
  obtained in many GR-based models. This type of result for RS-I model was
  also obtained in  \cite{avy}.

\bigskip

\centerline{\bf 4. Summary}

\smallskip

It is seen above that if phantom fluid dominates RS-I model of the homogeneous
and flat universe, where brane-tension $\lambda$ is positive,  universe
accelerates and is plagued with big-rip problem like most of the GR-based
models \cite{cal}. But, in case phantom fluid dominates RS-II model of the
universe (where $\lambda$ is negative) brane-gravity corrections make drastic changes in
the behaviour of phantom fluid. Unlike RS-I model, RS-II model of the phantom
universe expands till $\rho = 2 \lambda.$ It is found above that it explains
the present cosmic acceleration and exhibits deceleration after a finite time,
when energy density grows sufficiently. Phantom fluid is characterized by
violation of WEC. It is interesting to see that, in
the case of negative $\lambda$, brane gravity corrections suppress this
characteristic of phantom dark energy as energy density grows with time. When
$3|1 + w|/(1 + 3 |1 + w|) \le {\rho}/{\lambda} 
< (2 + 3|1 + w|)/(1 + 3 |1 + w|) $ WEC is not violated, but SEC is
violated. It shows that when $\rho$ satisfies this inequality, brane-gravity
corrections are so effective that phantom fluid behaves as quintessence. When
 $\rho$ grows more, brane-gravity corrections suppress even dark energy characteristics
 of the dominating fluid and neither SEC nor WEC are violated. As a result, phantom universe decelerates. Thus, here, it is
found that RS-II model based 
phantom universe accelerates in the present time and accelerating phase
continues upto a certain future time, but decelerates later till $\rho = 2 \lambda$. Moreover, it is
free from future-ingularity  as expansion of this model will stop when  $\rho
= 2 \lambda$. It happens so due to negativity of the 
brane-tension. 

\bigskip

\centerline{\bf  Acknowledgement}

\smallskip

Author is thankful to S.D.Odintsov for useful suggestions and comments.

\end{document}